# Exploring Students' perceptions of their learning experience and self-efficacy in physics online class with project-based learning: An interpretative phenomenological analysis

Mutmainna[1,2], Edi Istiyono[1], Haryanto[1], Beta Wulan Febriana[1,3]

[1]Graduate School of Educational Research and Evaluation, Universitas Negeri Yogyakarta, Sleman 55281, Indonesia
[2]Department of Physics Education, Universitas Sulawesi Barat, Majene 91412, Indonesia
[3]Department of Chemistry Education, Universitas Islam Indonesia, Sleman 55584, Indonesia

## Abstract

Project-Based Learning (PBL), recognized as an active learning strategy, has been linked to students' self-efficacy in prior studies, including those within Physics Education Research (PER). Meanwhile, technological advancements have significantly facilitated the optimization of diverse learning modes, including online learning. However, comprehensive investigations addressing questions such as *how students perceive PBL and their self efficacy, what forms of PBL design they prefer, and what benefits and challenges they encounter during its implementation* remain underexplored in existing literature. This study sought to uncover the experiences of ten students through longitudinal observation in an online PBL Physics class, focusing on its influence on their self-efficacy within a phenomenological study. Data were collected via semi-structured, in-depth interviews. Social Cognitive Theory, Bandura's Self-Efficacy Theory, and theoretical framework for the impact of project-based learning on educational outcomes were employed as guiding frameworks to shape the interpretation of students' experiences and perspectives. Through Interpretative Phenomenological Analysis (IPA), the researchers identified four primary themes representing the students' experience. There are (1) a variety of unique learning experiences based on students' perceptions about their character; (2) pivotal experience as a catalyst in the learning process; (3) student self-efficacy in Project-Based Learning; and (4) self-reflection based on prior experience. The findings contribute novel insights for educators in implementing PBL, particularly in online contexts, by highlighting strategies to accommodate diverse student personality characteristics. Moreover, the study provides helpful references for longitudinal research on the interplay between PBL and self-efficacy.

## I. INTRODUCTION

Educators in recent decades have experimented with a variety of learning strategies to find effective ways of teaching, such as flipped learning, project-based learning, problem-based learning, and cooperative learning. The goal is to welcome the paradigm shift from teacher-centered to student-centered learning [1]. Project-based learning (PBL), as one of the four learning strategies, has received a lot of attention in the field of education [2] The findings found that PBL significantly improved student learning outcomes, including academic achievement, motivation, self-efficacy and high-level thinking skills, increased student engagement in learning, and promoted the mastery of new knowledge [3-5]

Project-Based Learning (PBL), recognized as an active learning strategy, has been linked to students' self-efficacy in prior studies, including those within Physics Education Research (PER). Self-efficacy was introduced by a prominent Stanford psychologist, Albert Bandura, in his book, Self-efficacy: Toward a unifying theory of behavioral change, in 1997 [6]. Self-efficacy is defined as a person's confidence in his or her ability to complete a particular set of tasks or tasks in a particular field [7]. Furthermore, Bandura provides a detailed conceptual analysis and empirical review of how self-efficacy operates in conjunction with the sociocognition represented by Social Cognitive Theory (SCT) in influencing human actions, adaptations, and changes [8]. SCT itself is one of the theories of ideas that was also introduced by Albert Bandura in 1986 which is a study of many years of basic research using behaviorist and social learning frameworks [6]. These theories continue to have relevance in today's education field. For example, about adaptability, educating students to be able to adapt to future developments and face the complex real world has naturally become an important mission in education reform around the world.

Currently available study reports indicate that some research investigates the influence of PBL on self-efficacy in conjunction with other variables [1,9-13]. Furthermore, there is also research that only observes the application of PBL to self-efficacy [14-17]. The application of PBL and its relationship with other variables, for example, academic procrastination, science students' attitudes, learning, critical thinking, learning engagement, cognitive skills, and core competencies [13,18,19], summarize the findings of the application of PBL to student learning, affecting, for example, emotional attitude, thinking skills, cooperative ability [2], and combining PBL with other learning strategies, for example Flipped Learning [20].

Research focusing on the application of PBL and its relation to student self-efficacy shows that PBL has contributed positively to students' self-efficacy in the form of a statistically increased from before and after the implementation of PBL [14]. For example, solving Physics problems and student achievement [2,16,17,19], the survey results show a positive response with the implementation of PBL [10]. Furthermore, PBL is able to improve the ability to cooperate between students at different levels, and ages [1]. PBL as one of the active learning methods has been proven to reduce the self-efficacy gender gap in Physics learning [15] and increase engagement in the science environment [11]. Studies have proven the effectiveness of PBL when combined with other learning strategies [20]. Similar research also showed that self-efficacy was positively correlated with creative thinking attitudes and project-based learning value [12]. In addition to these positive findings, several reports were also obtained that provided recommendations in the implementation of PBL. For example, the findings of students' attitude scores towards PBL were lower than the scores of learning, critical thinking, and student engagement [13]. The findings are seen as having implications for how future PBL activities can be organized and how the curriculum can be developed to improve students' attitudes towards PBL.

Based on previous study, although there has been several empirical evidence of findings on the application of PBL specifically to self-efficacy, there are still several limitations in reporting the results. For example, in terms of research approaches, previous research was approached with a quantitative approach in experimental/quasi-experimental research [9,10,16,17,19,20] or by surveys [1,12-15], meta-analysis [2], by using questionnaire [1,12,13], by test [11,13,15], or by interview [10]. This indicates that the qualitative approach to exploring the application of PBL and its relationship with student self-efficacy remains limited.

The research approach in quantitative form reports the results of the study through statistical analysis based on responses to questionnaire/test. Despite previous research attempts to gather qualitative data through interviews, the exploration of students' perceptions of PBL implementation remains limited [10]. In addition, the study has not explored students' perceptions of their self-efficacy while participating in PBL. Furthermore, in general, previous research has not detailed the projects carried out in PBL, as well as the variability of observation time ranges, such as 3 weeks, 3 months, or even one year. Meanwhile, some other studies have not included the duration of the observations made. One of the research suggestions in previous studies was to explore the long-term impact of PBL implementation [18]

To address this gap in the literature, the authors used a qualitative phenomenological approach through Interpretative Phenomenological Analysis [21] to discover the lived experiences of students within four semesters (two years) of learning in an online class Physics to explore questions such as *how students perceive PBL and their self-efficacy, what forms of PBL design they prefer, and what benefits and challenges they encounter during PBL implementation*. A learning activity can seem very different to teachers and students. Improved alignment between teaching and learning requires knowledge of how students think and has proven to be very useful as a guide for effective teaching development [22].

The findings obtained in this study complement the results of previous studies. We believe that these findings can provide meaningful implications and knowledge contributions for teachers, pre-service teachers, and curriculum policy development institutions, especially in Physics Education and education in general. This study is guided by the following research question:

RQ1 : What is the student's perception of themselves and their learning experience in Project-Based Learning?

RQ2 : How is the Project-Based Learning design that students are interested in and how do students feel the benefits of the projects they are working on?

RQ3 : How do students perceive their self-efficacy based on their experience in doing assignments in Project-Based Learning? How does PBL give meaning to each participant's reflection at the end of learning?

## II. LITERATURE REVIEW

In this section, it will be explained conceptually related to PBL, Social Cognitive Theory (SCT) and Bandura's Self-Efficacy Theory. Social Cognitive Theory [23], Bandura's Self-Efficacy Theory [7], and theoretical framework for the impact of project-based learning on educational outcomes [2] were employed as guiding frameworks to shape the interpretation of students' experiences and perspectives. To explain the use of these three frameworks in the context of this research, this section is described.

Physics Education Research and Project-Based Learning

Studies related to curriculum and learning are one of the research topics in Physics Education Research (PER) [22-26]. In a study reported by [2]. Through the results of synthesis on 66 experimental research studies related to the application of PBL and its impact on student learning, a framework called the Theoretical Framework for the impact of Project-Based Learning on educational outcomes was obtained. In the framework, five main parts are explained, namely process, feature, ability, moderator, and learning effect. The process section explains the steps to implement PBL, while the feature section outlines what stimuli can be provided in the implementation of PBL to optimize learning. Furthermore, the ability section describes how the optimal implementation of PBL can influence students' abilities. The application of PBL may enhance creative thinking, cooperative ability, critical thinking, and computational thinking ability. In addition, PBL can train problem-solving, decision-making ability, problem raising ability, and comprehensive ability. PBL can also improve abilities in learning attitude, learning interest, learning motivation, and self-efficacy. In the moderator section, the variables that function as moderators related to the application of PBL to learning effects are described. These variables include subject category, course type, learning section, experimental period, experimental scale, and country region. Finally, in the learning effect section, in the form of thinking skills, academic achievement, and emotional attitude.

In this study, the use of a theoretical framework for the impact of project-based learning on educational outcomes is used as a lens to explore participants' perceptions in interpreting their experiences in PBL. For instance, how participants recall courses they identify as implementing PBL, the nature of their learning environment and approaches, as well as their perceptions of certain abilities they regard as influential, among other aspects.

Social Cognitive Theory **(SCT) and Self-Efficacy's Bandura**

Social Cognitive Theory (SCT) and Self-Efficacy theory were both introduced by Albert Bandura. The two theories are interrelated [6]. In a study written by [6], they present a framework that connects basic human capabilities in SCT. SCT generally explains the creation of reciprocal relationships between people, environment, and behavior [23]. Meanwhile, self-efficacy in general describes how experience, performance outcomes, vicarious experiences, social persuasion, and physiological feedback interact with each other and form self-efficacy in a person [7].

Stajkovic & Luthans [6] explain that basic human capabilities in SCT include symbolizing,

forethought, observational, self-regulatory, and self-reflective. Symbolizing describes the ability of humans to visually capture experiences and turn them into cognitive models that are used as a guide to deal with future events. Forethought refers to a person's ability to plan, anticipate consequences, and determine the purpose of the action. Observational explains how one observes the people or surrounding environment to make them a reference as well as reflect on the consequences of each chosen action. Self-regulation explains how a person controls themselves, sets standards internally, and evaluates the difference between those standards and the performance achieved, with the goal of continuously improving. Finally, self-reflection explains how a person reflects on their actions and determines, through perception, how strongly they believe that they can successfully complete a task in the future.

SCT and self-efficacy's Bandura in this study were used as a lens to explore how each participant interpreted their learning experience through the application of PBL. For example, what kind of experience makes them feel confident to be able to do the project assignment given by the teacher, what makes them feel challenged, how they relate to the experience they have had before, and so on. By exploring students' experiences in this manner, the aim is to provide teachers with insights for reflection in future teaching, particularly regarding the decision to implement PBL in online learning.

## III. METHODS

### A. Study Design

This research uses a qualitative-phenomenological approach [27,28] through Interpretative Phenomenological Analysis (IPA) [21]. Phenomenology is a qualitative approach that aims to find the essence of an individual's experience regarding a particular phenomenon [29], describes what consciousness feels and knows [30], and reveals the structure of meaning and how it is created [31]. In phenomenological research, phenomena are understood as mental events, events, or cognitive activities experienced by participants [29,32].

A study reported by [33] guides the author to gain detailed information about the complex phenomena of PBL in a Physics online class and to identify themes and theoretical structures that describe this process through IPA. IPA relies on three pillars. There are phenomenology (phenomenological philosophy), hermeneutics, and idiography [32]. Application of IPA In this study, it begins with the pillar of phenomenology applied by exploring each participant's response, with a focus on student live experience based on Martin Heidegger's views [34]. The emphasis in this section is on how the researcher seeks the interpretation/meaning of the learning experience by students through PBL applied by the teacher. Second, the pillar of hermeneutics was applied in a way that the researcher provides a way for each participant to interpret his or her experience during the implementation of PBL, and the researcher seeks to interpret the statement of experience conveyed by each participant (double hermeneutic) by applying *epoche* [35]. Third, the pillar of idiography was applied by seeing each participant as an individual who has uniqueness and uniqueness in interpreting their experience of undergoing PBL.

### B. Participants

This research involves the views of the university's Institutional Review Board as a guarantee of the implementation of research, especially protection for the participants involved in this research. Participant selection is carried out after the end of the active period of lecture activities and is carried out through the WhatsApp application. The researcher as well as the course instructor, distributed the online form into WhatsApp groups of students in two different batches. The screening process yielded up to 10 students who volunteered to participate in the research.

In selecting participants, the researcher adhered to the following criteria: (1) participants must confirm their willingness to voluntarily conduct interviews regarding their past learning experiences; (2) participants are active students of Physics education in the fourth or sixth semester who have programmed courses and actively participated in lectures through the application of PBL taught by the first author. A total of seven participants from the sixth semester were the primary focus of the study, while additional participants from the fourth semester were included to compare perceptions between the two classes at different levels. Based on these criteria, the author considers that the number of seven main participants in this study is still close to the number of participants recommended in the IPA, which is 3-6 people [21].

For participants who met both criteria, the researchers then sent private messages to them via WhatsApp to schedule interviews, and as part of the researchers' obligation to protect all collected participant data, we were guided by [36]. Detailed information related to the ten participants is presented in Table I.

TABLE I. Demographic Conditions of Participants

| Participant (pseudonym) | Semester | Course* | Perspective self-character |
|---|---|---|---|
| Gea | VI | 1 | Curious |
| Amy | IV | 2 | Dreamer |
| Zahra | VI | 1 | Introverted |
| Kanaya | VI | 1 | Humorous |
| Fidy | VI | 1 | Persistent |
| Ariqa | IV | 2 | Enthusiastic |
| Syita | IV | 2 | Hardworking |
| Indi | VI | 1 | Impatient |
| Lisa | VI | 1 | Optimistic |
| Cinta | VI | 1 | Socially reserv |

* (1) Development of Physics lesson plan
  (2) Physics learning strategy

The two classes chosen, namely the Development of Physics lesson plan documents and Physics learning strategy classes, consisted of 16 students and 21 students, respectively. In both classes, there are only two male students each. The first author distributed the form to recruit participants, but none of the four male students filled it out. This part is one of the limitations in this study.

## C. Research Setting

This research was the result of a longitudinal observation study of an online class of Physics students in the sixth semester, which was observed during four semesters of learning. The first author observed this class for two years, starting from semester III to semester VI. The first author taught five interrelated courses in the class, as presented in Table 2.

TABLE II. List of Courses

| Semester | Course Name | PBL Implementation Start from |
|---|---|---|
| III | Review of Physics Learning Curriculum | 13th meeting |
| IV | Physics Learning Strategies | 11th meeting |
| V | Physics Learning Assessment | 6th meeting |
| VI | - Development of Physics Lesson Plan<br>- High School Physics Experimental Design | 1st meeting |

Note:
The number of meetings in one semester of learning is 16 meetings (see attachment)

Table II in the PBL implementation column is the implementation of PBL that is implemented gradually by the researcher. There are several similarities between the five courses, namely:

1. All courses are carried out in the form of online learning using two modes, namely synchronous and asynchronous, through the Google Classroom platform. The first meeting determines the agenda (materials and activities) for each meeting. Each meeting includes material in the form of documents and videos, along with student activities and assignments. The appendix presents an example of one of the courses' displays.
2. The end-of-semester assessment consists of products resulting from the implemented project.
3. Each class is divided into groups based on specific criteria, with one member of each group serving as the coordinator to facilitate coordination with the course instructor.
4. Each student, even though they are in the same group, still has individual assignments, and each student's contribution has a direct impact on the group's assignment.
5. Each student began to compile and work on projects since the week of the PBL meeting. For instance, in the Review of Physics Learning Curriculum class, students initiate the end-of-semester project during the 13th meeting and complete it by the 16th meeting.

The difference lies in the level of difficulty of the projects being undertaken. The duration of the work aligns with the complexity of the assigned project. Given the difference in learning settings, the researcher intends to see the long-term impact of the implementation of PBL which is carried out in stages, on students' ability to optimize the output products of the course. Specifically, for courses that are in semester VI, the system is applied in the class as follows:

Students started working on the project during the first week of lectures.

Details of the topics for each meeting and the mini projects that must be completed each week have been conveyed since the first meeting.

The course content relates to the preparation of teaching materials (lesson plans) and the design of Physics experiments for senior high school. Classes are generally divided into three grade levels, corresponding to grades X through XII.

Each group consists of several students who divide the material content and experimental topics equally according to grade level.

The product for summative assessment at the end of the semester is to collect mini projects from start to finish.

The products produced from two courses are Physics Lesson Plan document, learning modules, and Physics experiments designs for grades X, XI, and XII in senior high school.

These two courses are offered to students one semester before they begin their school teaching practice, when they hold the status of preservice Physics teachers.

Semi-Structured Interview

Interviews are considered one of the key sources of information [37] and allow researchers to deeply explore a person's subjective attitudes and personal experiences [38]. This study employed semi-structured interviews lasting approximately 50–65 minutes using a virtual meeting application (Zoom). The first author acted as the primary interviewer during the data collection process. In conducting the interviews, the researcher provided an initial explanation and sought participants' permission to record the interview process. Recording began after participants gave their agreement. The file obtained is in the form of a .mp4 file but functions more as an audio file because generally the interviewer and the participant both agree to undergo the interview by only activating the microphone (without activating the camera), except for some participants.

During the interview process, we referred to the phenomenological approach by interpretative phenomenology [21]. The phenomenon in this study is interpreted as the awareness of participants participating in learning through the application of PBL. From this definition, the first author sought to explore how each participant uniquely perceives and interprets their experiences during the learning process (PBL), as well as their self-efficacy perceptions. During the interview process, the author explored the participants' experiences in the courses listed in Table 2, as well as in other courses that applied PBL. Table III presents the interview protocol we developed.

TABLE III. Sample Questions on the Interview Protocol

| No. | Theoretical Domain | Question Example |
|---|---|---|
| 1 | **Process of PBL** Exploration of the personal experience of PBL during learning | - According to the perception of your learning experience, what is the character of PBL?<br>- During your time as a student, what experiences did you have while learning with PBL?<br>- What kind of treatment do you think teachers do that can make them better understand the lecture material being taught? |
| 2 | **Ability in PBL** Explore what abilities in students can be influential in the implementation of PBL | What challenges did you feel during your time in college with PBL? What steps did you take to minimize that challenge? |
| 3 | **Self-efficacy** Seeking meaningful insights into how students view self-efficacy in the learning process. | How do you interpret the conditions of your environment, and how do you view your self-efficacy in working on projects during the lecture process? Have any previous experiences influenced your current view of self-efficacy? |

### D. Data analysis

After the interview data is obtained, the first author transcribes the interview recording file to obtain the interview transcript. In the transcript, the data for each participant has been anonymized. The data is then uploaded to a folder that can be accessed jointly by all members of the research team. Then the researchers began interpretive phenomenological analysis (IPA) of the data [21], which is inductive in nature. In this study, data analysis was carried out through the ATLAS.ti 25 [39] as the main application. In addition, some other applications, such as Microsoft Excel, are sometimes used as complements, especially in combining or separating certain themes.

In the data analysis, the researcher was guided by [32] and [21] through a study report by [33]. These steps consist of (1) reading many times (familiarizing the data); (2) making initial notes, (3) making an emergent theme; (4) creating a superordinate theme. Initial notes by researchers that include descriptive, linguistic, and conceptual comments [21]. In the initial phase, which involves familiarizing oneself with the data through repeated reading, researchers immerse themselves in the data to explore information based on the research questions. The next step in the process is to create initial notes. This stage is done by providing an interview transcript file with an extension .docx (Microsoft Word), presented in the form of a table, allowing the researcher to include exploratory comments. This exploratory comment is in the form of an interpretive statement of the researcher on the participant's statement that is felt to be important in the transcript [32].

The third step, creating an emergent theme, is done by the author by adding columns to the table in the transcript file to list the theme. The theme was born through the condensation of exploratory comments that had been made in the previous stage and were seen as interrelated. The theme is not expressed in the form of a question in the exploratory comment column but in the form of phrases or words. The last step is to create a superordinate theme. Superordinate themes are formed from the merging of several emergent themes into one larger theme. Superordinate themes are a collection of several emergent themes that have similar meanings [32].

## IV. FINDINGS

Three theoretical frameworks guide the presentation of the research findings. Social Cognitive Theory (Bandura, 1986), Bandura's Self-Efficacy Theory (Bandura, 1977), and theoretical framework for the impact of project-based learning on educational outcomes (Zhang & Ma, 2023) are used to answer research questions. The results section is organized based on four themes identified in the analysis process, which include (1) variety of unique learning experiences based on students' perception about their character; (2) pivotal experience as a catalyst in the learning process; (3) Student self-efficacy in Project-Based Learning; and (4) self-reflection based on prior experience. Here is a detailed explanation of these themes.

### 1. Theme 1: Variety of unique learning experiences based on students' perception about their character (RQ1)

Idiography, as one of the pillars in IPA, sees that each participant has uniqueness and uniqueness in interpreting their experience of undergoing PBL. In this section, participants' perceptions in interpreting the learning experience with PBL and how they view their own character will be presented by associating it with several keywords found in the participant's statement, such as leadership, positive beliefs, self-confidence, self-interest, barrier, and challenges. Additionally, the author explores how the participants' views are influenced by their individual characters when interpreting their experiences of learning through the models and strategies employed by their teachers.

Indi's experience relates to the leadership aspect, as she demonstrates impatience during PBL. Indi receives the assignment to take on the role of group coordinator. Indi's character, who tends to like things that are "instant," is conveyed through her statement, "PBL *gives me a challenge and gets me out of my comfort zone. Based on my character, who always wants to be instantaneous, I consider it heavy for project-based learning.*" However, Indi views her role as a leader as having a responsibility to set an example of good performance for her group members. This perspective can be seen in Indi's statement:

> Of course, it is very influential (the role of coordinator), Ma'am. For example, about the discipline of collecting tasks on time, the coordinator must first set an example for group members if he wants

> each group member to report progress (mini project) and be able to collect tasks on time.

As comparative data for participants from fourth-semester students related to leadership, Ariqa gives him a character as an *enthusiastic* figure when he becomes a coordinator. Ariqa provided a statement:

> From previous incidents (the character of the group members), I have coordinated continuously with my group to remind them that there are tasks like this and that but because they (group members) are not active, so I work only with one student. So, the assignments were the result of both of our thoughts, Ma'am.

The statement shows that Ariqa, as a coordinator, has problems in organizing his groupmates, especially in communication. When the task deadline approaches, the inactive group members casually work on the collected projects. Furthermore, the author traces the condition of the Ariqa group. For fourth-semester students, the implementation of PBL affords each group the freedom to determine the division of tasks among its members based on the group's internal agreement.

Furthermore, in the Ariqa group, there is no clear division of groups. This can be seen in the Ariqa question, "Only *one of my friends is active in the group, and there is no clear division of tasks in the group.*" In another part, Ariqa also stated, "*From this, I saw (how) the character of my friends. If it wasn't me trying to motivate them and start being active in the classroom, it would seem like it would just be like that".* In this statement, it appears that Ariqa, who interprets herself as enthusiastic, tries to encourage her classmates to always be actively involved in learning.

Next, related to how the perception of positive belief arises in the minds of participants when PBL is applied. For example, Zahra, who identifies as an introvert, chooses to share her experiences by expressing her habits related to her introverted nature. This was conveyed through a statement:

> The daily routine that I live with as an introvert. I am easily pessimistic, hesitant to express opinions, unconfident, and always thinks that things are difficult and unsolvable. So, I often lose before the match because my character is not confident.

From this statement, it appears that Zahra, who interprets herself as an introverted character, is very related to her perception of the confidence she has in doing something. However, when Zahra was asked about her perception of the implementation of learning through PBL strategies, Zahra stated:

> ...that I marked in PBL that the collaboration part of how to build cooperation with friends will have a very big impact, especially on me who is indeed difficult to socialize with and difficult to interact with friends. Even though I have known friends for a long time– it has been more than three years, but actually I only have two to three friends. So, PBL is very useful for the long term because it trains how to express opinions and how to unite opinions with friends when in the future it is needed.

Based on Zahra's statement, Zahra, with her introverted character, views that through PBL, she is given the opportunity to learn to socialize with her groupmates, including learning to express opinions, which is one of the things she did not dare to do before. Zahra's statement also shows how she struggles to interact with her classmates, which is evident from her limited number of close friends. Zahra believes that these abilities that are still her limitations are needed in the future, including when Zahra is in the world of work after becoming an alumna.

Next, Gea stated, "*Yes ma'am, so hmm... I then motivate myself also if, for example, this (PBL output product) is needed later during Teaching Assistance in the next semester. So, it has to be strong...*" Gea's statement indicates that she finds the routine of working on projects in PBL to be burdensome. However, Gea sees it as motivation and believes that the learning experience he gained will be needed when Gea participates in Teaching Assistance (field experience), so Gea must be strong in undergoing the learning process.

In line with Zahra and Gea's statement, Amy stated, *"For the manufacture of products, maybe the goal is so that the theory we get can be put into practice immediately."* Based on this statement, it appears that Amy interprets the application of PBL to make the theories taught by teachers easier to understand through practice. Another opinion related to positive faith was also expressed by Amy when asked to tell her experience when the course teacher asked to interview one of the Physics teachers at school. Amy asserts, "The teacher possesses extensive experience. *Indirectly, the information provided is important for us in the future.*" Amy's statement shows that through one of the assignments in PBL, she gained experience that helped her to look positively at the people she interacted with, one of which was through an interview assignment.

Furthermore, the participant's statement related to self-confidence starts from Gea, who gives a statement:

> It influences Ma'am (PBL's duties on Gea's confidence) … but at the same time it also comes at the same time as doubting, for example muttering whether what I am doing is right or not, or whether it is in accordance with what the lecturer should have requested.

Based on this statement, the assignment of tasks in PBL has influenced Gea's confidence. However, in carrying out this task, the question always arises whether the product he works on meets the criteria given by the teacher or not. Furthermore, in relation to the level of difficulty of the given project, Gea gave a statement, "*Usually, if for example, we (Gea and group friends) are working on project assignment and have reached the stage of dizziness (feeling difficult to do the task), we feel that either the task is difficult or we are slow and late to do the task.*" Based on this statement, Gea felt doubtful when facing a situation where the tasks collected were not optimal. Gea considers that this feeling of doubt may be due to his limited abilities, or it may also be triggered by time management factors in doing tasks.

Meanwhile, Zahra, regarding her confidence during PBL, gave a statement:

> ...Because my character is not confident. I feel very overwhelmed (understanding the material or doing assignments) all this time because I don't dare to express my opinion. I was always afraid to ask because I thought that my questions should not be wrong and I was embarrassed if my questions were easy for other friends.

Based on this statement, it appears that Zahra lacks confidence, which affects her learning process, such as difficulties in expressing opinions or asking teachers about lecture material that they do not understand. For the fourth-semester participants, Amy also expressed her feelings related to self-confidence by giving a statement, "*When there is a project task, in my head there is no idea. I sometimes say whether I can or not.*" This statement hints at how Amy felt confused and clueless when she was first given a project assignment, as well as showing how her confidence at the beginning of the lecture felt doubtful about her abilities.

In undergoing PBL, self-interested participants seem to have their own role. Gea, through his statement, said, "*I prefer the PBL show, which is dominant in its own creativity.*" Gea's statement indicates that she prefers to work on projects that require creativity and individual responsibility. Still related to self-interest, Gea, in another statement, gave an explanation,

> But it seems that your push is also strong, ma'am. And that's the reason I remember the most when designing the learning module, I tried to optimize how to do it so that the results were good. For example, if it's work, honestly, I'm lazy. But because it requires creativity and confidence, finally I often spend as well. In the end, (the module) feels the most optimal.

Based on Gea's statement, stimulated self-interest arises because he considers that designing modules provides freedom to be creative independently. Ultimately, Gea believes that she has effectively optimized her products.

In line with Gea, Amy gave a statement," For *me, for the timing, it's probably better, ma'am* because I like the media." Amy's statement shows that her love for designing learning media helps her to better organize her time when working on projects related to learning media.

Furthermore, regarding barriers, Zahra stated, *"In general, my experience with this PBL learning model is that I have a lot of difficulty in working in groups or collaborating with friends."* According to Zahra's statement, it shows that one of Zahra's main obstacles in PBL is working in a group. Furthermore, Zahra explained:

> I experience three different feelings during group work when working on a project. The first is a condition that I think is safe. All members of the group can be invited to discuss. So, there are several friends that I feel comfortable discussing with to be able to work on a project. Second, there is only one student who is active in the group, while the other members dominantly leave it to the group leader only. So only one person is responsible. Third, the group of two to three students who I think are smart. However, this project could not be completed on time because it was difficult for these three students to unite opinions. Each with his or her own desire.

Zahra, who experienced obstacles in group work, revealed that during PBL, she was able to identify various group characters that interacted with her. Of the three conditions described, only one condition makes Zahra feel safe, which is when interacting with certain friends who are considered to be able to collaborate well during discussions to work on projects. Although Zahra briefly mentioned the barriers she faced within her group of friends, Gea provided a more detailed explanation of the obstacles that affected her personally.

Gea stated," The result (the output product) can be maximized actually, just because I am not able to divide my time even better." This statement shows that the barrier that Gea feels comes from himself.

In line with Gea's statement, Amy, through her statement, said," While working on projects, I often produce suboptimal results due to poor time management. I procrastinate too many times. Too much time spent, especially on social media." Amy's statement suggests that her excessive use of social media leads to poor time management. The barrier in the form of time management is also felt by Syita, who labels herself as hardworking, with the statement, "From previous experiences, what made us (Shia and her group) less confident was actually from ourselves, especially in the timing." Syita's statement showed that the timing that became a barrier in working on the project also impacted her confidence.

Regarding the challenge, Gea stated, "Honestly, in the first session (lecture) I had trouble developing learning tools." Gea's statement shows that the challenge faced by Gea is to develop learning tools. In the learning setting applied by the first author, the learning tool is the first product, and the learning module is the second product. Meanwhile, a participant from the fourth semester, Amy stated, "My experience made me feel a little challenged because in the Physics curriculum Study course, where we (students) had to make observations to school. It was the first time for me to do an interview with a teacher." The statement indicates that Amy views the new experience as a challenge she must overcome while undergoing PBL.

In the next part, the author explores participants' experiences with offline and online learning, as well as their perceptions of conventional learning and PBL, based on each participant's character. Students from semester VI participate in offline lectures during their first year, transitioning to online lectures in subsequent years, including semester VI. As for the participants of the fourth semester, they underwent the first year of online lectures from the beginning of the lecture period.

Statements related to learning modes, e.g., through Lisa's statement, *"If I didn't have time at that time (think of the idea of doing a virtual interview) because of the period of adjusting to online learning. So, using Zoom (virtual meeting application) is still lacking."* Based on the statement, Lisa, who had to switch from face-to-face learning mode for the first year to online learning, required adaptation, including the use of several applications. One of them is the use of virtual meetings when getting a mini project assignment, namely conducting an interview. Next, Gea stated:

> For offline, because it is carried out directly, they are more enthusiastic in working on projects. Especially if the project is in a group, in working on the project, they also encourage each other. So, there is more enthusiasm in doing it, and the results (external products) are also faster to see (the progress).

Gea's statement showed that Gea interpreted his experience by feeling more suitable to work on group project tasks offline. Some of the benefits obtained are mutual encouragement between group members and the progress of output products that are more visible. Contrary to Gea's perception, Zahra gave a statement," For *me personally, I prefer online learning. The main reason is because it goes back to my character that likes a safe place to study for me.*" Zahra's statement indicates that, as an introvert, she feels that studying in a private setting is safer and allows her to better understand the lecture material and complete assignments effectively. However, Zahra also stated in another statement,

> For online learning, I consider the challenge to be bigger. The first is networking (technical obstacles), and the second is that if this PBL is applied to work in a group lesson (obtained) yesterday (the previous time), it is difficult to communicate. We (Zahra and her group) can communicate on the day of the course (synchronously). However, (in asynchronous mode) we find it difficult to be active at the same time, even though the time has been determined for a discussion together. None of them were timely for various reasons, so we couldn't have a discussion.

Based on Zahra's statement, it appears that although Zahra basically likes online learning because she feels safer in learning, she admits to experiencing technical problems related to networking and difficulty coordinating with fellow group members, especially in asynchronous mode. For Gea, online learning is reflected in his statement, "For online learning, because we (students) work alone, the encouragement of laziness is also big." For Gea, learning that is done individually causes greater laziness than doing tasks offline. For comparison, the participant of the fourth semester, Syita, gave a statement that was in line with Zahra:

> ... During the most difficult online course project, the network (internet). Then in terms of getting information, I also have problems because sometimes I look for information on the internet, but it is difficult to understand.

Based on this statement, Syita faces several obstacles in online learning, including technical issues related to the internet connection and challenges in understanding and processing information obtained from the internet. However, when participants were asked if there was a difference in the sense of responsibility in doing tasks between offline learning and online learning, they confirmed that there was no difference in the two learning modes. This can be seen in Zahra's statement, "For *me, it's the same (online and offline learning).*"

As for the participants' perceptions of the relationship they felt between conventional learning and PBL, when Amy was asked how she interpreted her experience when she was assigned to interview a Physics teacher directly in a mini project at PBL compared to the conventional learning that she generally felt before, Amy stated, "*It will be much different (conventional learning and PBL). For example, in interview skills..."* Amy's statement shows that she interprets her experience through PBL to feel much different than through conventional learning.

2. **Theme 2: Pivotal experience as a catalyst in the learning process (RQ2)**

In this section, the author explores the participants' perceptions related to PBL. This includes how they interpret PBL, what rules are memorable for each participant in undergoing

PBL, and how they uncover ways to learn. In addition, the author observed participant testimonials and identified variables that may have influenced the participants during their experience with PBL.

The description of the findings began with how each participant interpreted PBL. This is, for example, stated by Gea by giving the keyword, namely "progress." Based on that keyword, it appears that Gea interprets the main characteristic of PBL as the periodic progress of tasks/projects. Gea further revealed, *"... Projects are created within a certain period of time and then there are sessions where students are trained to be consistent with progress at all times."* Gea's statement demonstrates the need for consistency in the progress of project tasks under PBL. Meanwhile, Zahra tends to express her perception by examining whether the courses she went through have applied PBL. This is evident in his statement, *"… in your class (first author), the six stages (PBL) exist, but there was a project in the previous course last semester, and not all of the stages exist."* This statement shows that Zahra, who has taken the Physics Learning Strategies class, took advantage of the opportunity in the PBL lecture to observe whether the PBL syntax is applied in full or only partially in a particular course.

As a comparison of participants from semester IV, Ariqa interpreted PBL through the statement, "PBL *produces products. What I remember most about PBL is the project or product, which is expected to make these students better understand the materials.*" Based on Ariqa's statement, it appears that Ariqa interprets the existence of projects carried out in PBL as a means that is believed to help him to better understand theoretical lecture material, which is then practiced through the project he is working on.

Furthermore, some of the rules that are memorable for each participant in undergoing PBL and how the participants find ways to learn are obtained from several statements, including: (1) agreement in PBL; (2) preferred PBL design; (3) implementation of a time schedule for one semester; (4) monitoring of assignment progress by teachers; and (5) a collection of Lecture Materials. Agreement in PBL for example can be seen in Gea's statement, *"The methods taught are like there must be progress every week, so we are required to study harder and finally be able to understand."* Based on Gea's statement, it appears that the emphasis on reporting on project progress every week makes Gea feel like he has to study harder, which ultimately helps her understand the learning material.

In line with Gea's statement, Zahra said in her statement, *"It doesn't matter (reporting progress every week), it makes us better. We who were relaxed became more agile."* Zahra's statement showed that through reporting the progress of the project that was carried out every week, Zahra felt that she got benefits related to managing the time to work on tasks.

Regarding the PBL plan that the participants liked, in another statement, Gea stated," I *prefer PBL's designs that are more demanding in their own creativity.*" The statement hinted that the type of project that Gea likes is one that gives him space for his own creativity (without intervention from his groupmates). In PBL applied to the participants, the space to create independently is facilitated through the preparation of learning modules with different material content for each individual in session II of the lecture in the Development of Physics lesson plan class. In contrast to Gea, Kanaya said in his statement:

> ...The course that prepares the lesson plan, because it is explained in detail from the beginning to the end. There is an example of a document. So, the impression is that we as students can understand well. I like the way the teacher teaches because it is explained until the students understand. It doesn't have to be chasing time. Once the students understand the subject, they will move on to the next subject.

While Fidy blocked, "*In my opinion, you really appreciated our comfort in learning activities, because it was rare for anyone to ask us such questions (Fidy and his classmates). And 80% of teachers don't ask (students about the form of learning that students prefer)*". Based on this statement, it appears that Fidy positions himself as a student who considers it important for a teacher to ask his students if the strategies applied are in accordance with the way students learn based on their character. When Fidy was asked about the role or importance of teachers in evaluating the learning strategies applied, Fidy stated, "*I think it plays a very important role, because my motivation to learn has increased. And for me, the level of comfort and how the student's learning conditions or situations are decisive.*" Based on Fidy's follow-up statement, it appears that Fidy likes the PBL design in which there is a session where the teacher asks for evaluation from students directly.

Furthermore, the participant's statement related to the implementation of the time schedule for students for 1 semester, for example, through Fidy's statement:

> Because there (Google Classroom) the deadline for each assignment has been described, and the material that is being or has been discussed is also listed. Students can listen back to the lecture materials. And for me it makes me disciplined – *ok, this hour or this meeting should have been collected assignments about...*- I muttered like that.

Fidy's statement shows that there is an implementation of a time schedule for one semester, which can help Fidy to be more disciplined in doing assignments. In another statement, Lisa gave a statement:

> So, from the positive value, for me it was taught how to manage time, how to really make a good schedule, and how to target the final results that will be achieved. Then from the negative value, when other courses also demand to complete certain assignments, the assignments at the same time often make me complain a little. So, I thought that the task might be a lot, but compared to the positive side, it trained me to manage my time well so that I achieved my study goals well.

Based on Lisa's statement, the implementation of a time schedule with a target project that must be achieved every week by every student has positive and negative sides for her. However, Lisa acknowledges that the positive impact is significant, particularly in fostering her ability to manage her time effectively.

In addition to the implementation of a time schedule for one semester, teachers also routinely monitor the progress of assignments in synchronous classes that have been done by students. Several statements obtained are related to this, for example, through Gea's statement, "*... Because several times the teacher did comment (on my assignment). Usually also what is commented on in my friends' assignments (by the teacher) also happens to be that there is also a deficiency in my assignment, so I can take (references) from there.*" Gea's statement suggests that teachers use progress monitoring to rectify independently completed assignments.

Furthermore, Amy, a participant from semester IV, stated, "*I like the teachers who control the progress of the project we (students) are working on.*" Amy's statement indicated that she liked it if there were sessions in learning where teachers monitored the progress of student assignments. In line with Amy, Zahra stated:

> What I really like is when the teacher monitors students based on the progress of the project and checks the products one by one. Why I like that stage is because I know the extent of my work, what I understand, and whether the execution is correct or still wrong. So, at the time of checking, I found out that, - *okay, I was low on this part, I was wrong in this part.*

Zahra's statement indicates her preference for teachers monitoring individual or group tasks, as it allows her to independently evaluate her work. Meanwhile, Indi stated, "*In my opinion, if it is per progress or there is a schedule, the possibility of plagiarism is small because it is checked one by one by the teacher.*" Based on Indi's statement, checking the progress of assignments by teachers can minimize the occurrence of plagiarism in student assignments. This is in line with Fidy's statement, "*For the content of the assignment, it is completely different, and we (students) are really looking for references and then looking for our own ideas.*" Fidy's statement indicates that the projects developed by each student in each group require them to seek out references and ideas individually.

In contrast to the previous statements, the monitoring of task progress through the provision of comments on students' assignments, when discussed collectively by the instructor in synchronous online learning, Cinta who described herself as socially reserved, stated *"When the instructor comments on my work, for me there is both a pleasant and an unpleasant feeling"* This statement indicates that when Cinta's work was commented on by the instructor, there were conditions that made her feel uncomfortable. Another statement from Cinta revealed:

> It was my first time making a product and the first time my work was commented on. At that time, I wanted to be told which part was wrong, but that day I felt (*the sentence had been interrupted for a long time, a sign of hesitation in choosing a word - there was a whisper of reflex...-eeehhh, well, I don't know how to pronounce it*). In essence, at that time I felt like I was cornered, so I was also embarrassed by my classmates.

Based on this statement, it appears that Cinta once felt that she had an unpleasant experience with the implementation of PBL applied by teachers. Furthermore, statements related to the form of lecture material files in the form of teacher explanation videos, in addition to files in the form of reading materials in .pdf, for example, can be seen in Zahra's statement, *"I like online learning accompanied by videos that I can play back because for me to be able to understand the material explained by the teacher, it cannot be just once, but it needs to be listened to repeatedly."* This statement shows that Zahra needs to play the video repeatedly to understand the material well, so she likes the PBL design that comes with a video explanation of it. In line with Zahra, Kanaya stated, *"The video is very useful because if I have a piece of material that I missed, I can play it back, and the video can be used as a benchmark for the preparation of the next assignment."* For Kanaya, the existence of videos is considered very useful, both for repeating the material that has been passed as a reference and for compiling project assignments given by the teacher.

Furthermore, in addition to exploring what the participants' experiences were, the author also collected several testimonials from them during PBL. For example, Cinta stated, *"If I am asked directly to make the product, I will know better what is called a Lesson Plan, which is called a syllabus, what its use, and so on."* The statement indicates that through the implementation of PBL, Cinta can directly observe the sample document in question and learn to distinguish various learning tools through product creation. Next, Fidy gave a statement, *"I am happy because at PBL I am trained to think creatively to generate ideas. Then many abilities are involved, such as improving technological capabilities when looking for references or when working on projects."* Based on this statement, it appears that for Fidy, PBL provides many benefits, including training Fidy to think creatively and demanding complex abilities.

Examples of products worked on by participants, for example through, Lisa's statement, *"Previously, the learning media I made was about the theory of gas kinetics by utilizing used waste."* Lisa's statement showed an example of a product that has been developed for learning media, investigating the theory of gas kinetics by utilizing used materials from the surrounding environment.

The participants' testimonials regarding the benefits felt by participants after entering the observation period at school as preservice Physics teachers were found in several statements. For example, through Indi's statement, *"Previously, there was a product to make modules, a product that we produced from courses on campus. It turns out that there are no learning modules available at school, so it is as if we have come to bring something new to the school. We design ourselves. And for me it was very helpful."* Based on Indi's statement, it appears that the modules generated from the course can be used to teach at school during practical teaching activities, and Indi finds the material very helpful.

Furthermore, Fidy stated, *"I was assigned to class XII and my products are applied at school. I am thrilled. It turns out that the product I made can be applied in school."* Fidy's statement showed that there was a sense of happiness in him when he realized that the practicum design product he made, which is a product of the output of the course on campus, could be used at school.

There are several participant statements that show how PBL is also connected to student problem-solving. For example, through Gea's statement, "*If I feel lazy to work on a project, for example, I come to a friend's house to motivate each other or I can also go for a walk for a while... The term is to find a solution to create a good mood first and then start working on the project again.*" Gea's statement shows that in undergoing PBL, there are conditions where Gea's motivation decreases and he tries to find solutions in various ways. One of the ways Gea does this is to go to his friend's house with the aim of motivating each other or to go for a walk to make Gea's mood conducive to returning to work on the project.

Another statement came from Kanaya, who lost the task file during the seconds of the collection deadline. Kanaya stated, "*I still try to work again, ma'am. That night I cried and felt emotional. But the next day I immediately worked on it for two nights. I was grateful that my task was completed before the deadline. It turned out that I could still get through it. I thought it wouldn't be possible, but it turned out that it could still be.*" The sentence shows how Kanaya chose a solution to rework a project that should have been completed. Kanaya chose to vent her emotions first and then immediately tried to solve her problem. Although he had doubts about her abilities, Kanaya finally realized that he had successfully passed the challenge well.

### 3. Theme 3: Student Self-Efficacy in Project-Based Learning (RQ3)

This section describes participant statements that pertain to student self-efficacy in Project-Based Learning. This theme encompasses several emerging aspects, specifically the self-ratings of each participant regarding their self-efficacy and their perceptions of student self-efficacy while engaging in PBL.

Participants provide statements about their self-efficacy by choosing a rating on a scale of 1 to 4. For example, through Gea's statement, "*I'm on a scale of three, because I feel like I can still get to the max but haven't been able to get to that point.*" This statement shows that Gea feels that the project that has been worked on can still be further optimized, and he realizes that he has not reached his maximum potential.

In a subsequent statement, Cinta stated, "*I am on a scale of 2 because I feel that in the preparation of my product, I still need to learn a lot.*" Cinta sees that he still needs to learn a lot, so he chooses a scale of 2. Zahra gave a statement:

> My current self-efficacy is rated at 2.5 on the scale. At present, I remain far from achieving a high level of self-efficacy, as I continue to experience considerable pessimism. Prior to taking this course, I perceived my self-efficacy to be at zero. Thus, reaching 2.5 represents a notable improvement.

Zahra's statement shows how at the beginning of the lecture she interpreted her self-efficacy to be at zero. And when she chose the 2.5 scale, Zahra interpreted it as a significant improvement. As for Fidy, through his statement, "*I chose four. Because from PBL activities there are many new lessons that I have learned so that my confidence and confidence in myself have increased greatly.*" Based on Fidy's statement, she chose the highest scale because she felt that she had learned many new lessons and that made her confidence increase significantly.

Next, related to how the view of student self-efficacy while undergoing PBL, for example, is found in Gea's statement, "*Wow, it turns out that I can get to this stage, and maybe in the future I can do more by trying a little more to get more maximum results.*" Gea's statement shows his appreciation for his achievements after going through a series of activities in PBL, and that she has plans to improve his business in the future so that he can achieve more optimal results from what has been achieved.

Meanwhile, Zahra, through her statement, said, "*Previously, I had difficulty coordinating my friends to work and informing friends who were not active to work. The courage came spontaneously because it was a deadline.*" Based on this statement, it appears that the deadline for collecting projects in PBL has encouraged Zahra to improve her leadership skills in coordinating her groupmates even though she considers herself to have an introverted personality. Lisa expressed, "I have gained a wealth of experience. *So, there is starting to be confidence and being able to compete with friends. Not competing for anything but learning more than what was expected from the previous time.*" Lisa felt that her confidence increased after seeing her achievements in PBL. He felt that he could achieve better learning outcomes compared to the previous semesters, and with that he realized that he was able to compete with his peers academically. Lisa emphasized that academic competition is about being able to see herself always achieve more academic achievements than what is targeted.

### 4. Theme 4: Self-Reflection based on prior experience (RQ3)

In this section, the author describes the statements of the participants related to reflection based on the experiences that have been passed, including the experiences gained through the implementation of PBL. For example, through Zahra's statement, "*I became more confident in expressing my opinion. That's what I consider to be a positive impact on my self-efficacy which was previously very low.*" The challenges that Zahra went through in PBL have made her feel more confident, especially in expressing her opinion.

Regarding plagiarism, Indi shared her experience when she admitted that she had done an assignment by copying her classmate's work. This is evident in her statement," Previously*, it was because I wanted to make progress (task progress), I just copied my friend's work. And it made me not understand the material because I was just copying. And finally, I understood that doing it ourselves turns out to make it easier for us to understand the material well*". Based on this statement, Indi interpreted her experience of copying a friend's work, making her unable to understand the material. And from that experience, he realized that reporting progress on a regular basis gave him the opportunity to work on tasks based on her own thoughts and avoid plagiarism.

Meanwhile, Kanaya gave a statement, "*Honestly, yesterday I was a little overwhelmed, but it was good for me because I felt like I was being disciplined. At first, I wasn't sure but over time I came to be sure.*" Kanya's statement shows his acknowledgment of feeling uncomfortable (feeling burdened), but with it she realizes that there is a change for the better (becoming more disciplined) in doing her tasks. As for Shiita, she stated, "*Very positive experience of the assignment given. At the time I became a teacher. Even though it was only in front of my friends, I could feel that being a teacher was so difficult.*" Syita expressed her experience as a teacher in a simulation of learning activities in one of the courses through PBL. From her experience, she learned that being a teacher is not an easy thing.

The participant's statement related to awareness to learn can be found in Indi's statement

> All knowledge turns out to be useful. Nothing we end up saying –*it's useless for me to learn this and that. turned out not to be implemented*. I also finally realized from this class that, in addition to learning the material, the class activities helped shape my character to be more mature. Previously, I had always relied solely on my own values. Now I am grateful to be more open when it comes to business problems with the results or appreciation of others with our efforts.

Based on Indi's statement, she tries to interpret all forms of their experience while undergoing PBL. Indi feels that her activities during the course are not just learning material. However, Indi feels the process of becoming more mature. Previously, Indi saw that his achievements in the academic field only relied on grades. Indi realized that her efforts in learning also had an important meaning for her.

*Summary of Results*

The description of the research findings indicates that several keywords can effectively summarize the results of the analysis of the respondents' experience perceptions. The way each participant labels their characters gives a unique color to how they interpret their respective experiences. In general, participants interpret the implementation of PBL from a variety of perspectives. For instance, the study revealed that participants understood the importance of responsibility in leadership. Additionally, they believe that the learning experience from PBL will be beneficial for their future. At the beginning of the PBL implementation, participants generally felt doubtful about their ability to complete the project, especially due to the learning that occurred in the online mode. Technical obstacles such as the internet network became a challenge for the participants in coordinating with group friends. Furthermore, in undergoing PBL, there are several factors that function as catalysts for the participants in optimizing the learning process. One such factor includes participants' interest in the project content, as well as the PBL design imposed by the teacher.

In this study, some catalyst factors are regular monitoring of students' project progress. The presence of monitoring is considered a forum for students to be able to independently evaluate the project tasks that have been done, make improvements, and look at examples contextually so that the study of lecture material is listened to through theory and practiced directly. PBL is interpreted as reporting the progress of tasks periodically, but it is also interpreted as a process of building character, including practicing discipline and perspective that leads to maturity in looking at things happening around participants through a reflection process. The initial sense of doubt slowly grew and improved. Although there are still many things that can be optimized, in general, the participants realized that they learned by doing, interacted with their classmates, and applied learning by doing. The transition from offline to online learning with learning by doing, followed by the accumulation of these experiences through reflection, fosters self-efficacy and changes perspectives that grow from before.

## V. DISCUSSION

This study is a longitudinal observation in an online Physics class that focuses on courses that provide the basics of pedagogy one semester before semester VI students practice teaching at school. After undergoing PBL for four semesters, this study aims to explore how participants interpret their experiences and findings that can complement those of previous studies. The findings are believed to have important implications, especially in Physics learning for preservice Physics teachers. Based on the findings of the research, the result will be discussed in this section.

### 1. RQ1: What is the student's perception of themselves and their learning experience in Project-Based Learning?

To answer RQ1 related to the first theme, Variety of unique learning experiences based on how students' perception about their character. Students' perceptions of themselves and their interpretations of their learning experiences in PBL reveal diverse meanings based on their individual characteristics, which are influenced by various aspects. In general, it appears that the participants are trying to adapt to PBL. For instance, one participant prefers an instant process, but as coordinator, she learned that she had to set an example for her friends to collect tasks on time. Other participants, who see themselves as introverts, do not give up on their position as coordinators even though communication with their groupmates sometimes encounters certain obstacles. PBL encourages students to be able to develop various abilities, including comprehensive abilities [2]

Regarding online learning, in general, the responsibility in doing assignments is interpreted as no different between online and offline learning. For participants with characters other than introverts, they generally tend to prefer offline learning. The data shows that the learning mode chosen by the participants is related to their respective characters. These findings provide insight for teachers or preservice teachers to be able to understand how students' character plays a role in determining their perspective on learning, especially through the application of PBL in online classes.

As for their view on the application of PBL, in general, the participants indicated that they felt that they had gained a lot of meaningful experiences that were more diverse than conventional learning (in the form of classes that only focused on theoretical discussions). For instance, they believe that PBL aids in the development of interview skills and other practical skills, which conventional learning fails to provide. This is contrary to the findings of previous research, for example by [13], who found that the score of students' attitude towards PBL was lower compared to the scores of learning, critical thinking, and student involvement in learning.

2. **RQ2 : How is the Project-Based Learning design that students are interested in and how do students feel the benefits of the projects they are working on?**

To answer the RQ2 question related to the second theme. Pivotal experience as a catalyst in the learning process, several findings obtained include how the participants interpreted PBL itself. The PBL designed in the context of this study includes periodic progress reporting for projects carried out over a semester. Previously, the participants had been given treatment in the form of PBL learning, but the implementation period of PBL was different. This consideration is carried out as an effort to form an environment for the implementation of PBL gradually for students. Students often struggle to adapt to the PBL model due to their familiarity with conventional, teacher-centered learning methods [40]. In addition, the initial phase of the PBL model, which involves project planning and scheduling, requires the facilitation of proficient teachers to engage students in generating project ideas that are relevant to the learning material [4].

In the implementation of PBL, most of the participants interpreted PBL as a learning process that is synonymous with the word "progress." Participants feel experiences that focus on developing discipline, such as reporting progress periodically, managing time well, receiving feedback from teachers regarding projects that have been done, and finding solutions to various challenges encountered during the PBL process. Participants also hold the belief that PBL positively influences future learning experiences.

The application of longitudinal PBL encourages a participant to re-evaluate whether the learning steps implemented by the teacher have covered all stages of PBL or only partially. These findings show how a student is able to evaluate the implementation of PBL as a follow-up to the theoretical knowledge that has been acquired by students, especially those related to Physics learning strategies. Thus, the result indicates that the participant has reached the metacognitive learning stage. For a preservice Physics teacher, observing and evaluating the learning syntax shows achievements in the design thinking phase [41]

Some key discoveries about designing PBL that create valuable experiences for participants include the project's content, the way instructions are given for group and individual tasks, having a clear semester schedule, regular evaluations of the teacher's support, consistent feedback from lecturers, and teachers keeping track of assignments. In addition, adjusting teaching materials to the online learning format also plays an important role. These findings, which complement previous research, indicate that the application of PBL needs to be understood qualitatively in addition to being evaluated quantitatively. PBL is thought to function better for the laboratory class compared to the theory class [2]. From the findings obtained from this study, the authors argue that PBL, applied even in theoretical classes, can also function well.

Regular evaluation of the treatment provided is a specific consideration for certain participants. The author interpreted the question of how involved students felt in the learning system as meaning participants felt actively involved, including in designing their learning. Furthermore, the implementation of PBL was also seen to have a positive impact on other skills, such as problem-solving and creative thinking. These findings align with reported studies [2], [12].

Regarding task monitoring, participants received and interpreted it positively. Analysis and evaluation of project outcomes is essential, especially in the context of teaching, where teachers play a role in guiding students to interpret and conclude project outcomes using appropriate concepts and presentations [9]. However, this section needs to acknowledge certain findings. Teachers must deliver monitoring assignments that involve commenting on the products students have worked on carefully. Such behavior This is important because inappropriate comments can affect students' confidence. In addition to building engagement with all students in the class, a personal approach is also an important part for a teacher to pay attention to, including recognizing the student's character well.

The implementation of online classes that provide reading materials and explainer videos accessible without restriction is perceived by participants as facilitating their learning and understanding of the subject matter. This unrestricted access across space and time enables them to study independently. Furthermore, the use of course products beyond the classroom is regarded by participants as a valuable experience. According to them, the outputs of the course serve as tangible evidence of their work and have proven applicable during teaching practicum activities as preservice Physics teachers.

### 3. RQ3: How do students perceive their self-efficacy based on their experience in doing assignments in Project-Based Learning? How does PBL give meaning to each participant's reflection at the end of learning?

To answer the RQ3 question related to the third and fourth theme, Student Self-Efficacy in Project-Based Learning and Self-Reflection based on prior experience. The observation carried out for two years (4 semesters) raises questions for the author about the extent of the application of PBL to the self-efficacy of the participants. The findings in this section include self-ratings conducted by participants regarding their self-efficacy, as well as their views on the role of self-efficacy in participating in PBL.

The findings of self-rating related to self-efficacy in participants showed mixed results, ranging from the lowest scale to the highest scale based on the scale range given by the researcher. Two findings emerged from participants who selected a low self-rating scale. First, their PBL experience led them to believe that they still have much to learn. Second, participants felt that their self-efficacy at the beginning of learning was at a scale of zero, so giving a rating on the middle scale was considered an extraordinary achievement. On the other hand, participants who gave their assessments close to the optimal scale felt that the product produced could actually still be improved more than what had been done. Participants who chose the highest rating considered that the experience they had gone through made them aware of the many new skills they had mastered. Additionally, they achieved a product that they consider to be the best possible result of their efforts. The implementation of PBL strategies offers many advantages for students, including influencing goal orientation, increasing curiosity, increasing engagement, and encouraging the mastery of new knowledge [5].

Furthermore, the findings of this section reveal variations in participants' reflections. These include perceived ability before and after learning, the impact of PBL on self-discipline, reflections of students in their role as preservice Physics teachers, and the way PBL fosters participants' maturity in perceiving themselves as learners. The results are consistent with the three theoretical frameworks used to examine the participants' experiences [2,7,23].

The participants' reflections on their abilities before and after learning, on how they evaluate their actions and believe they can succeed in subsequent tasks, their reflections as preservice Physics teachers, as well as their reflections related to discipline and the role of PBL as part of the process that fosters maturity, are in line with Self-efficacy theory [7] and Social Cognitive Theory [23]. These aspects include performance outcomes, vicarious experience, and physiological feedback within Self-efficacy theory, as well as observational, self-regulatory, and self-reflection aspects within SCT. The perceived ability developed during PBL is consistent with the findings reported by [2].

We acknowledge the limitations of our study. Although we aim not to generalize our findings in other contexts, we want to confirm that the results reported in this study solely reflect the data obtained from participants who were voluntarily involved in this research. Research participants who only involve one type of gender (female students) are one of the limitations in this study. Furthermore, a number of findings on how PBL leads to students' design thinking abilities as preservice Physics teachers can be studied in the next research. In addition, the existence of learning videos published through the YouTube platform in this study provides an opportunity to develop research focused on Learning Analytics (LA) and Educational Data Mining (EDM). These two areas can serve as focal points for future research.

## VI. CONCLUSION

The differing meanings of "character" for each participant in this study have created unique learning experiences for them, which in turn affects how they interpret their experiences during the learning process. A variety of factors contribute to how participants feel interested in actively engaging in learning and the benefits of what they have been doing. Although PBL has been applied in longitudinal observations, the data show that it does not necessarily have a significant effect on the meaning of self-efficacy in each participant. However, it seems clear that previous experiences played an important role in shaping how participants viewed their own self-judgment.

The participants stated that there was a relationship between PBL and the self-efficacy they felt. Finally, reflection is the final stage of PBL syntax, resulting in variations in the form of reflection found in the research data. The findings of the research are believed to be able to make a meaningful contribution for teachers, especially Physics teachers, in choosing PBL as a learning strategy to be applied in the classroom. These findings are not only supported by quantitative evidence from previous studies but also by the qualitative approach we report in this study.

**AUTHOR CONTRIBUTIONS:**
Conceptualization, M.M.; methodology, M.M; formal analysis, M.M., B.W.F; data curation, M.M., B.W.F; writing—original draft preparation, M.M., B.W.F; writing—review and editing, M.M., B.W.F; supervision, E.I., H.H. All authors have read and agreed to the published version of the manuscript.

**ACKNOWLEDGEMENT:**
The first author would like to thank the Ministry of Higher Education, Science, and Technology (KEMDIKTISAINTEK), the Center for Higher Education Funding and Assessment (PPAPT), Indonesian Endowment Fund for Education (LPDP) of the Republic of Indonesia, and Indonesian Education Scholarship (BPI) for funding in the first author's study doctoral and this research.

**CONFLICTS OF INTEREST:**
The authors declare no conflict of interest

# APPENDIX

Example of a display of one of the PBL implementation courses through online classes on the Google Classroom platform on Development of Physics Lesson Plan class

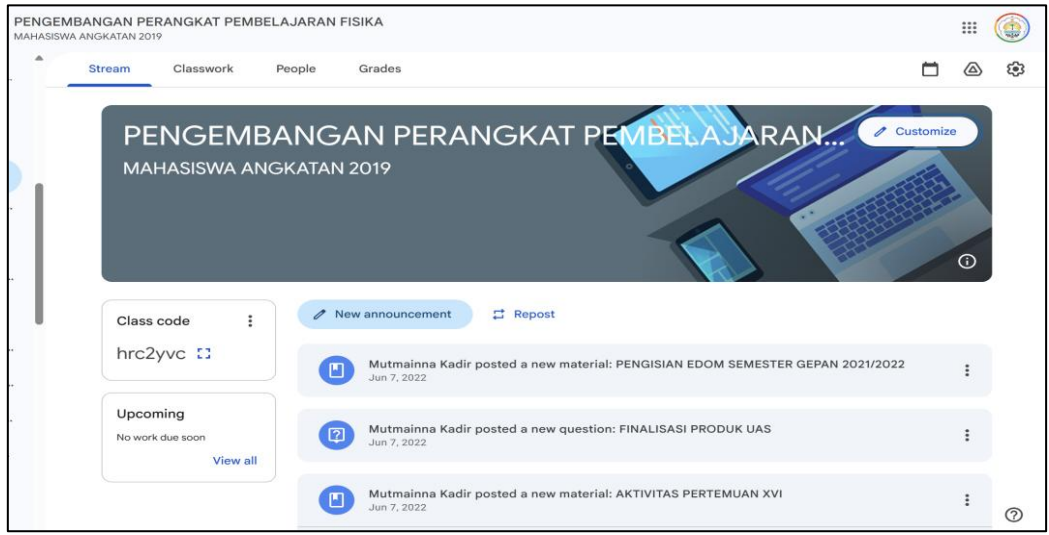

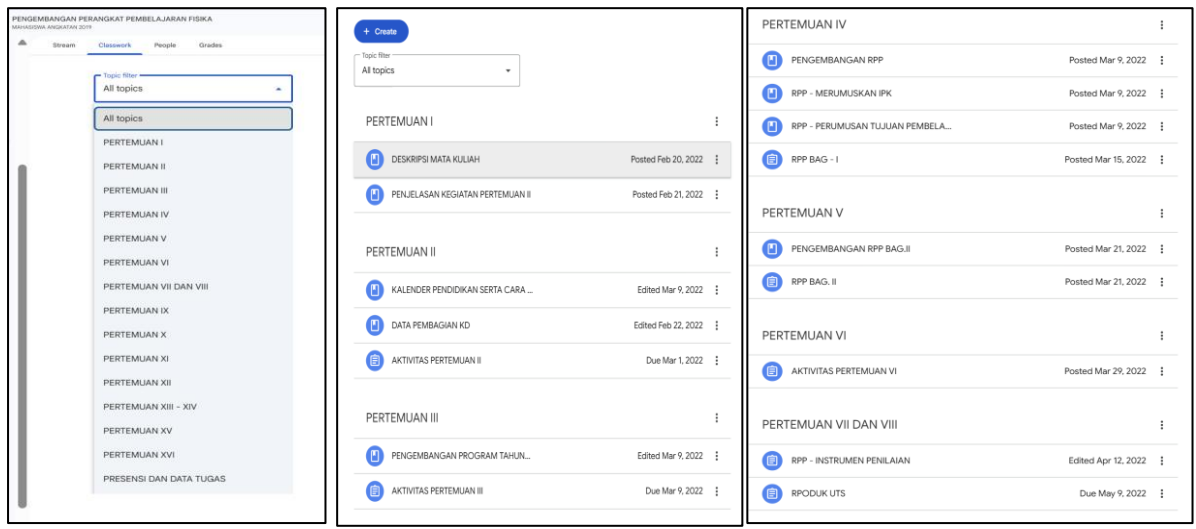

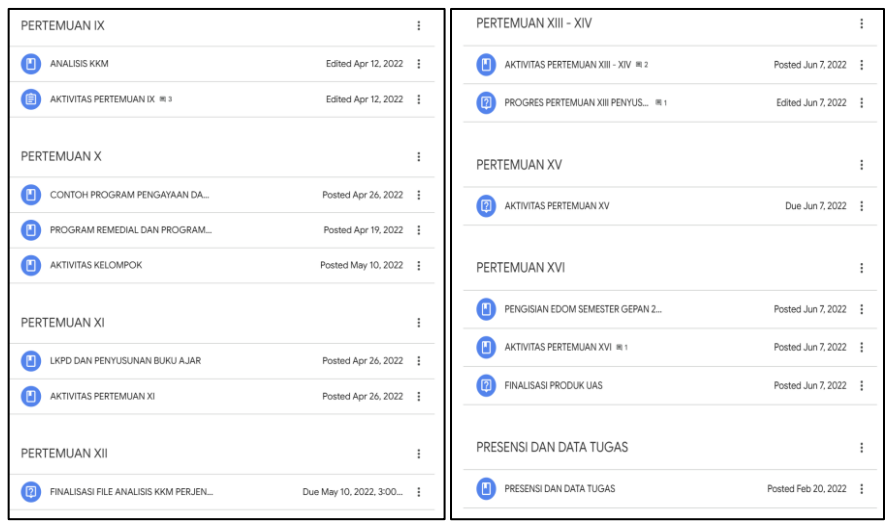